\documentclass{mn2e}

\usepackage{amsfonts}
\usepackage{amsmath}
\usepackage{graphicx}

\newcommand{\avg}[1]{\ensuremath{\langle \,#1\, \rangle}}

\newcommand{\be}{\begin{equation}}
\newcommand{\ee}{\end{equation}}

\title[Excursion set with correlated steps]
      {One step beyond:  The excursion set approach with correlated steps}

\author[M. Musso \& R. K. Sheth]
{Marcello Musso$^{1,2}$\thanks{E-mail: marcello.musso@uclouvain.be, sheth@ictp.it} 
 \& Ravi K. Sheth$^{1,3}$\\
 $^1$ The Abdus Salam International Center for Theoretical Physics, 
      Strada Costiera 11, 34151 Trieste, Italy\\
 $^2$ CP3, Universit\'e de Louvain, 2 Chemin du Cyclotron, 
      1348 Louvain-la-Neuve, Belgium \\
 $^3$ Center for Particle Cosmology, University of Pennsylvania, 
      209 S. 33rd St., Philadelphia, PA 19104, USA}
\begin{document}
\pagerange{\pageref{firstpage}--\pageref{lastpage}}

\maketitle 

\label{firstpage}

\begin{abstract}
We provide a simple formula that accurately approximates the first 
crossing distribution of barriers having a wide variety of shapes, 
by random walks with a wide range of correlations between steps.  
Special cases of it are useful for estimating halo abundances, evolution, 
and bias, as well as the nonlinear counts in cells distribution.  
We discuss how it can be extended to allow for the dependence of the 
barrier on quantities other than overdensity, 
to construct an excursion set model for peaks, 
and to show why assembly and scale dependent bias are generic 
even at the linear level.  
\end{abstract}

\begin{keywords}
large-scale structure of Universe
\end{keywords}

\section{Introduction}
In hierarchical clustering models, to estimate cluster abundances at any 
given time one must estimate the abundance of sufficiently overdense 
regions in the initial conditions (Press \& Schechter 1974).  
The problem is to find those regions in the initial conditions that are 
sufficiently overdense on a given smoothing scale, but not on a larger 
scale.  The framework for not double-counting smaller overdense 
regions that are embedded in larger ones is known as the Excursion Set 
approach (Epstein 1983; Bond et al. 1991; Lacey \& Cole 1993; Sheth 1998).  

In this approach, one looks at the overdensity around any given random 
point in space as a function of smoothing scale. The resulting curve 
resembles a random walk, whose height tends to zero on very large 
smoothing scales.  In this overdensity versus scale plane, the 
critical density for collapse defines another curve, which we will call 
the barrier.  The double-counting problem is solved by asking for the 
largest smoothing scale on which the walk first crosses the barrier.  
It is fairly straightforward to solve this problem numerically, by 
direct Monte-Carlo simulation of the path integrals (Bond et al. 1991). 
This is particularly simple if the steps in the walk are independent, 
but one can also include correlations between the steps, whose nature 
depends on the underlying fluctuation field (i.e., for a Gaussian field, 
on the power spectrum) and the form of the smoothing filter.  

When the steps are independent, exact solutions for constant 
(Bond et al. 1991) or linear (Sheth 1998) barriers are known.  
Exact solutions for more general barrier shapes are not known, but 
good analytic approximations are available (Sheth \& Tormen 2002; 
Lam \& Sheth 2009).  In the appropriate units, these solutions are 
self-similar, i.e.~independent of the form of the power spectrum 
(of course, they depend strongly on the barrier shape).  
However, an exact solution of the first crossing problem for correlated 
steps is still unknown.
The main goal of the present work is to provide a simple formula which 
works for a wide variety of barrier shapes, smoothing filters, and power 
spectra.  This is done in Section~\ref{main} -- our main result is 
equation~(\ref{sfs}), and it is explicitly not self-similar.  
Section~\ref{extend} describes a number of extensions of this 
calculation -- having to do with walks conditioned to pass through a 
certain point in a certain way, or with barriers whose height depends 
on hidden variables.  
A final section summarizes our results, indicating how we expect 
our work to be used when fitting to the halo abundances which, 
recent simulations indicate, are not quite self-similar.

\section{First crossing distribution with correlated steps}\label{main}
In what follows, we will assume that the underlying fluctuation field 
is Gaussian.  We comment on non-Gaussian field in the Discussion section.  
In hierarchical models, the variance of the fluctuation field is a 
monotonic function of the smoothing scale (the exact relation depending 
on the shape of the power spectrum and the 
smoothing window.)  Therefore we can use the terms smoothing scale and 
variance interchangeably, and we will use $s$ to denote the variance.  

Let $B(s)$ denote the height of the barrier on `scale' $s$, 
and $p(\delta,s)$ the probability that the walk has height $\delta$ 
on this scale.  We will assume that $\langle\delta\rangle\equiv 0$, 
so that $s\equiv \langle\delta^2\rangle$.  We would like to write 
down the probability $f(s)$ that $\delta < B(S)$ for all $S<s$ and 
$\delta > B(s)$ at $s$.  When the steps in the walk are uncorrelated 
the two conditions separate, simplifying the analysis.  
For correlated steps, this simplicity is lost.  

\subsection{The completely correlated limit}
Recently Paranjape et al. (2012) have argued that there is considerable 
virtue in thinking of the limiting case in which the steps in the walk 
are completely correlated.  In this case, if the walk had height 
$\delta_0$ on scale $S_0$, then it has height
 $\delta = \delta_0 \sqrt{s/S_0}$ 
on scale $s$, so $\delta$ is a smooth monotonic function of $s$.
Therefore, if $\delta$ first exceeds $B(s)$ on scale $s$, it was 
certainly below $B(s)$ on all $S<s$, and one need not account for this 
requirement explicitly.  Hence, the first crossing distribution is just 
\begin{align}
 f_{cc}(s) &=
 \frac{\partial}{\partial s} \int_{B(s)}^\infty {\rm d}\delta\,p(\delta,s)
 \nonumber\\
 &= -\frac{\exp(-B^2(s)/2s)}{\sqrt{2\pi}}\,
      \frac{{\rm d}B(s)/\sqrt{s}}{{\rm d}s}
      \quad {\rm for}\ s < S_{\rm min},
 \label{fcc}
\end{align}
where $S_{\rm min}$ is the scale on which $B(s)/\sqrt{s}$ is minimum 
(in many cases of interest $S_{\rm min}=\infty$).  
Note that this relates the shape of $f(s)$ to that of $p(\delta,s)$ 
on the same scale.  

Paranjape et al. (2012) show that, despite the strong assumption about 
the deterministic smoothness of the walks, this expression provides a 
very good description of the first crossing distribution (at small $s$) 
even when the steps are not completely correlated.  Physically, this is 
because if one thinks of real walks as stochastic zig-zags superimposed 
on smooth completely correlated trajectories, at small $s$ a walk has 
not fluctuated enough to depart significantly from its deterministic 
counterpart.  
One might thus expect that there is danger of double counting trajectories 
with two or more crossings only when $s$ becomes large.

In cosmology the large $s$ regime is not nearly as interesting as the 
small.  The discussion above suggests that it will be useful to construct 
an expansion in terms of the number of times walks cross the barrier.  
The first step in this program is to assume 
that no walks double-cross, but the actual correlation structure scatters their crossing scale around the completely correlated prediction. Accounting for this fact alone should allow to estimate $f(s)$ with a greater regime of accuracy than equation~\eqref{fcc}.
Accounting for one earlier crossing should be even more accurate, and so on. 

\subsection{A bivariate approximation for the strongly correlated regime}
To proceed, we assume that when steps are strongly correlated one can 
replace the requirement that $\delta(S)<B(S)$ for all $S<s$, which is a 
condition on all the steps in the walk prior to $s$, with the milder 
requirement that $\delta(s-\Delta s)<B(s-\Delta s)$ for $\Delta s\to 0$ -- 
a condition on the one preceding step. Because now the walk values on the 
two scales are independent, the analysis is more involved than when the 
steps were completely correlated, but the increase in complexity is 
relatively minor because we only require a bivariate distribution.  

For small $\Delta s$ we can expand both $\delta$ and $B$ in a Taylor 
series.  The condition on the walk height at the previous step means 
that 
 $\delta(s) - \Delta s\,\partial \delta/\partial s 
  < B(s) - \Delta s\,\partial B/\partial s$.
If we use primes to denote derivatives with respect to $s$, 
then the first crossing distribution of interest is given by the 
fraction of walks which have
\begin{displaymath}
 B(s)\le\delta\le B(s) + \Delta s\,(\delta' - B')
 \quad {\rm and}\quad  
 \delta' \ge B'(s).
\end{displaymath}
That is to say, 
\begin{align}
 f(s)\,{\rm d}s &= \lim_{\Delta s\to 0}
 \int_{B'}^\infty {\rm d}\delta'\, \int_{B(s)}^{B(s) + \Delta s\,(\delta'-B')}
    {\rm d}\delta\,p(\delta,\delta') 
 \nonumber\\
 &= \Delta s\, p(B,s) \int_{B'}^\infty {\rm d}\delta'\,p(\delta'|B)\, 
                     (\delta' - B').
 \label{limfs}
\end{align}

For a Gaussian field $p(B,s)$ is Gaussian, and the conditional 
distribution in the integral is also Gaussian, with shifted mean 
and reduced variance.
If we define 
\begin{equation}
 \gamma^2\equiv \frac{\langle\delta \delta'\rangle^2}
                     {\langle\delta^2\rangle \langle \delta'^2\rangle}
 \qquad{\rm and}\qquad
 \Gamma^2 = \frac{\gamma^2}{1-\gamma^2}\,,
 \label{gammas}
\end{equation}
the mean of $p(\delta'|B)$ is $\avg{\delta'|B} = \gamma\,B
\avg{\delta'^2}^{1/2}/\avg{\delta^2}^{1/2}$ and the variance is 
$\avg{\delta'^2}(1 - \gamma^2)$.  
(Our notation was set by the fact that, for Gaussian smoothing filters, 
$\gamma$ equals the quantity spectral quantity 
 $\sigma_1^2/\sigma_0\sigma_2$ which Bardeen et al. 1986 called $\gamma$.  
For tophat filters, the integrals over the power spectrum which define 
our $\gamma$ are given by Paranjape et al. 2011.)  Note that 
$\avg{\delta\delta'} = 1/2$, and thus $\avg{\delta'|B} =B/2s$.

Before we evaluate the integral, note that the completely correlated 
approximation corresponds to the limit in which $\gamma=1$ and 
$p(\delta'|B)$ becomes a delta function centered on $B/2s$.
The integral then yields
 $B/2s - B'= -\sqrt{s} \partial (B/\sqrt{s})/\partial s$, and
the resulting expression is consistent with equation~\eqref{fcc}.  
Notice that our analysis has indeed extended the completely correlated 
solution by replacing the delta function with a Gaussian whose width 
depends explicitly on the underlying power spectrum.  

In the generic case, one still has an integral over a single Gaussian distribution. If we define
\begin{displaymath}
 x \equiv \frac{\delta'-B'}{\langle\delta'^2\rangle^{1/2}},\ \ 
 \beta(s) \equiv \frac{B(s)}{\sqrt{s}} \ \ {\rm and}\  
 \beta_*(s) \equiv -\beta(s)\frac{\partial\ln\beta(s)}{\partial \ln\sqrt{s}}
\end{displaymath}
(the reason for our choice of sign for $\beta_*$ will become clear later),
then equation~\eqref{limfs} gives
\begin{align}
 sf(s) 
  &= \frac{{\rm e}^{-\beta^2(s)/2}}{2\gamma\,\sqrt{2\pi}} 
       \int_0^\infty \!\!{\rm d}x\,x\,
     \frac{{\rm e}^{-(x - \gamma\,\beta_*(s))^2/2(1-\gamma^2)}}
            {\sqrt{2\pi(1-\gamma^2)}} .
\label{likepeaks}
\end{align}
Evaluating the integral yields
\begin{equation}
 sf(s) = \frac{{\rm e}^{-\beta^2(s)/2}}{2\,\sqrt{2\pi}}\,
         \beta_*\left[\frac{1 + {\rm erf}(\Gamma\beta_*/\sqrt{2})}{2} 
       + \frac{{\rm e}^{-\Gamma^2\beta_*^2/2}}{\sqrt{2\pi}\Gamma\beta_*}\right]
\label{sfs}
\end{equation}
(recall $\beta_*$ depends on $s$); this is our main result.

Comparison with equation~(\ref{fcc}) shows that the term in square 
brackets above represents the correction to the completely correlated 
solution.  There are two points to be made here.  
First, this correction term depends on $\Gamma$, indicating that the 
shape of the first crossing distribution depends explicitly on the 
form of the underlying power spectrum.  In this respect, walks with 
correlated steps are fundamentally different from walks with uncorrelated 
steps.  
Second, when $\beta_*\gg 1$ then this term tends to unity so the 
first crossing distribution reduces to that for completely correlated 
walks.  To see what large $\beta_*$ implies, it is useful to consider 
some special cases.  

\begin{figure}
 \centering
 \includegraphics[width=\hsize]{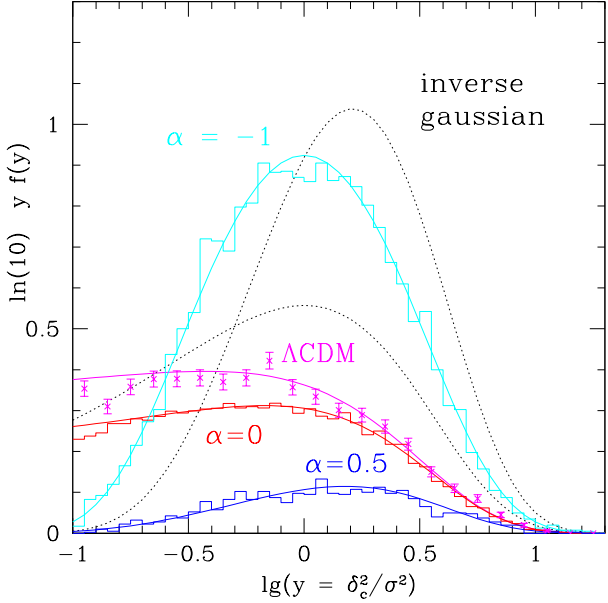}
 \caption{Distribution of the scale $y=\delta_c^2/s$ on which walks 
          first cross the barrier $\delta_c(1 + \alpha s/\delta_c^2)$.
          Histograms show numerical Monte Carlo results for Gaussian 
          smoothing of $P(k)\propto k^{-1.2}$, for which $\Gamma^2 = 9/10$; 
          top to bottom are for barriers with $\alpha = -1,0$ and $0.5$.  
          For comparison, the two dotted curves show the corresponding 
          distributions for $\alpha=0$ and $-1$ when steps are uncorrelated.
          Smooth curves show Eq.~(\ref{sfs}) with $\Gamma^2=9/10$ and 
          the appropriate values of $\alpha$; the agreement indicates that 
          our formula works well for a wide variety of barrier shapes.  
          Symbols with error bars show results for a constant barrier 
          ($\alpha=0$) and tophat smoothing of a $\Lambda$CDM $P(k)$.  
          This shows that, in contrast to when steps are uncorrelated, 
          the first crossing distribution does indeed depend (weakly) 
          on $P(k)$.  The solid curve shows Eq.~(\ref{sfsconstant}) 
          with $\gamma=1/2$ for which $\Gamma^2=1/3$; the agreement shows 
          that our formula works well for a wide range of $\Gamma$.  }
 \label{compareMC}
\end{figure}

\subsection{When the barrier has constant height}
For a constant barrier, $B(s) = \delta_c$ and 
$\beta_*(s) = \delta_c/\sqrt{s}$.  The latter is conventionally 
denoted as $\nu$, so that $\beta_*\gg 1$ corresponds to large $\nu$.  
In terms of $\nu$, equation~\eqref{sfs} is 
\begin{equation}
 \nu f(\nu) = \frac{\nu\,{\rm e}^{-\nu^2/2}}{\sqrt{2\pi}}\,
         \left[\frac{1 + {\rm erf}(\Gamma\nu/\sqrt{2})}{2} 
        + \frac{{\rm e}^{-\Gamma^2\nu^2/2}}{\sqrt{2\pi}\Gamma\nu}\right].
\label{sfsconstant}
\end{equation}
In this form, it is clear that $\Gamma^{-1}$ acts as the $\nu$-scale below which the correction to the completely correlated distribution becomes important.  (Note however that our $\Gamma$ is not the same parameter defined by Peacock \& Heavens 1990 and used in Paranjape et al.~2012.)

Figure~\ref{compareMC} compares this formula to distributions 
generated via Monte-Carlo simulation of the walks (following 
Bond et al. 1991).  We present results for two rather different power spectra and smoothing filters, 
with two different values of $\Gamma$.  
Our first choice is Gaussian smoothing of $P(k)\propto k^{n}$ with 
$n=-1.2$, for which $\Gamma^2 = (n+3)/2 = 9/10$.  
Our second is tophat smoothing of a $\Lambda$CDM power spectrum; 
in this case $\gamma$ itself depends on scale, with $\gamma\approx 1/2$ (and hence $\Gamma\approx 1/3$) on the scales where $\nu\approx 1$.  
The figure shows that equation~(\ref{sfsconstant}), with $\Gamma^2 = 9/10$ 
and $1/3$ respectively, provides an excellent approximation to the 
Monte-Carlo distributions.  Note in particular that, for $\Lambda$CDM, 
ignoring the scale dependence of $\Gamma$ (by simply using $1/3$ on all 
scales) works very well.  This demonstrates that 
Eq.~(\ref{sfsconstant}) is a simple and accurate approximation for 
the CDM family of models.  

\subsection{Linear barriers} 
For linear barriers $B(s) = \delta_c (1 + \alpha s/\delta_c^2)$, 
making $\beta(s) = \nu + \alpha/\nu$ and $\beta_* = \nu - \alpha/\nu$.
Figure~\ref{compareMC} shows results for $\alpha = -1$ and $1/2$, both for 
Gaussian smoothing of $P(k)\propto k^{-1.2}$; 
Equation~(\ref{sfs}) with $\Gamma^2 = 9/10$ works very well.  
Note in particular that our formula is quite different from the 
Inverse Gaussian distribution associated with uncorrelated steps.

\section{Extensions}\label{extend}

The analysis above has been so simple, and its results so accurate, 
that it is interesting and natural to extend it to a variety of other 
problems, some of which we outline below.  

\subsection{Halo bias}\label{bias}
The excursion set approach is often used to quantify the correlation 
between halo abundances and their environment.  This is done by 
computing the ratio of $f(s|\delta_0,S_0)$, the first crossing 
distribution subject to the additional constraint that the walks 
passed through some $\delta_0$ on some large scale $S_0$ before first 
crossing $\delta_c$ on scale $s\gg S_0$, to $f(s)$. 

Motivated by the previous Section, we now set
\begin{equation}
 f(s|\delta_0,S_0)\,{\rm d}s = 
 \int_{B'}^\infty \!\!{\rm d}\delta'\!
 \int_{B(s)}^{B(s) + \Delta s\,(\delta'-B')}
    \!\!\!\!\!\!{\rm d}\delta\,p(\delta,\delta'|\delta_0,S_0) 
\end{equation}
in the limit where $\Delta s\to 0$.  This will give  
\begin{equation}
 f(s|\delta_0,S_0) = p(B|\delta_0,S_0)
 \int_{B'}^\infty \!{\rm d} \delta' p(\delta'|B,\delta_0) \,(\delta'-B')\,,
\end{equation}
which is very similar to equation~\eqref{limfs}, only with modified 
mean values and variances. The Gaussian outside the integral has mean 
$\avg{\delta|\delta_0} = \delta_0 \avg{\delta\delta_0}/S_0$ and variance 
$s (1 - \xi^2)$, where $\xi\equiv \avg{\delta\delta_0}/\sqrt{sS_0}$.  
The one inside has mean
 $\langle\delta'|B,\delta_0\rangle = (\mu_B B + \mu_0 \delta_0)/(1 - \xi^2)$, 
where 
 $\mu_B = [\langle\delta\delta'\rangle - 
  \langle\delta'\delta_0\rangle\langle\delta\delta_0\rangle/S]/s$
and 
 $\mu_0 = [\langle\delta_0\delta'\rangle - 
   \langle\delta\delta'\rangle\langle\delta\delta_0\rangle/s]/S_0$.  
The first Gaussian obeys the scaling assumed by Paranjape et al.~(2012), who showed it was a good approximation to their conditional Monte Carlo distributions. Therefore, it is interesting to see if this scaling also holds for the integral.

If $\langle\delta_0\delta'\rangle\approx 0$ (or, more carefully stated, 
if the term containing this quantity is smaller than the other) then 
 $\mu_B = \langle\delta\delta'\rangle/s$ and 
 $\mu_0 = -\langle\delta\delta'\rangle 
           \langle\delta\delta_0\rangle/sS_0$, 
so 
$\langle\delta'|B,\delta_0\rangle
 = \mu_B (B - \langle\delta|\delta_0\rangle)$.  
Since this is the same rescaling of $B$ as for the first Gaussian, in 
this approximation $f(s|\delta_0,S_0)$ follows the scaling 
with $\delta_0$ that was assumed by Paranjape et al. (2012).  
Therefore, the conclusions of Paranjape \& Sheth (2012) about the 
difference between real and Fourier space bias factors also hold for our 
calculation.  In particular, the Fourier space bias $b_1$ will be simply 
given by differentiating $\ln\, sf(s)$ with respect to $B(0)$, whereas the 
real space cross correlation will carry an additional factor of 
$\langle\delta\delta_0\rangle/S_0$.  
For a constant barrier 
 $b_1= \partial\ln \nu f(\nu)/\partial\delta_c$ 
yields
\begin{equation}
 b_1 = \frac{\nu^2-1}{\delta_c} + \frac{1}{\delta_c}
         \left[1 + \frac{1 + {\rm erf}(\Gamma\nu/\sqrt{2})}{2} 
               \frac{\sqrt{2\pi}\Gamma\nu}{{\rm e}^{-\Gamma^2\nu^2/2}}\right]^{-1}.
\label{bias1}
\end{equation}
The first term on the right hand side is the bias associated with 
completely correlated walks.  The correction term is negligible for 
$\nu\gg 1$, and it tends to $1/\delta_c$ when $\nu\ll 1$.  

The expression above assumes that $\langle\delta'\delta_0\rangle$ is 
negligible.  This is reasonable when $S_0\ll s$;  e.g. for Gaussian 
smoothing filters,
 $\langle\delta'\delta_0\rangle = \sigma^2_{1\times}/2\sigma^2_1$ falls 
to zero faster than $\langle\delta\delta_0\rangle$.  But at intermediate 
scales it is not, and $f(s|\delta_0,S_0)$ will have additional 
dependence on $\delta_0$.  
In Musso, Paranjape \& Sheth (2012) we show that this will generically 
introduce scale dependence into the Fourier space bias factors, even at 
the linear level.

\subsection{Conditional mass functions}\label{fsS}
The excursion set theory can also be used to compute halo progenitor 
mass functions and merger rates (Lacey \& Cole 1993). For this, we 
need the joint distribution $f(s,b,S,B)$ of walks that first cross the 
barrier $B$ on scale $S$, and first cross the barrier $b>B$ on scale 
$s>S$. Dividing by $f(S,B)$, which we already have, will give the 
conditional distribution $f(s,b|S,B)$.  The most natural way to 
estimate it is to set 
\begin{align}
 f(s,b&|S,B) \approx \frac{1}{\Delta s}
 \int_{B'}^\infty {\rm d}\Delta' \int_{B(S)}^{B(S) + \Delta S\,(\Delta'-B')} d\Delta
 \nonumber\\
 & \qquad \times\int_{b'}^\infty \!{\rm d}\delta' 
     \int_{b(s)}^{b(s) + \Delta s\,(\delta'-b')} \!\!{\rm d}\delta\,
  \frac{p(\delta,\delta',\Delta,\Delta') }{f(S)dS}
     \nonumber\\
&= \frac{p(B)}{f(S)}\int_{B'}^\infty \!{\rm d}\Delta'\,(\Delta'-B')\,p(\Delta'|B) 
     \,p(b|B,\Delta') \nonumber\\
 & \qquad \times \int_{b'}^\infty {\rm d}\delta'\,(\delta'-b')\,
            p(\delta'|b,B,\Delta')\nonumber\\
&= \frac{p(B)}{f(S)}\int_{B'}^\infty {\rm d}\Delta'\, (\Delta'-B')\,
       p(\Delta'|B)
  \,f(s|B,\Delta').
 \label{sfsS}
\end{align}
(Strictly speaking, we should adjust the limits on the integrals 
over $\Delta$ to ensure that $\Delta < b$.)  
We have written the final expression in a suggestive form, to show that 
the left hand side should be thought of as a weighted average over first 
crossing distributions, each with its own value of $\Delta'$.  

In the limit where the scales are very different, $s\gg S$, it should 
be a good approximation to ignore the 
$\langle\delta\Delta'\rangle$, $\langle\delta'\Delta\rangle$ and 
$\langle\delta'\Delta'\rangle$ correlations.  This makes  
$b - \langle b|B,\Delta'\rangle = b - \xi (B - G\Delta')/(1-G^2)$
where $G$ is the same quantity as $\gamma$, but on scale $S$. 
This means there is some dependence on $\Delta'$.  
Note that if there were no correlation with $\Delta'$ (in effect $G\to 0$)  
then
the conditional distribution would have the same form as the 
unconditional one, except that
 $b\to b - B S_\times/S$, and $s\to s -  S (S_\times/S)^2$,
where we have defined $S_\times \equiv \langle\delta\Delta\rangle$.  
This is similar to the rescaling for sharp-$k$ filtering, for which 
$S_\times=S$.  
However, for Gaussian smoothing of a power law $P(k)$, 
$G=\gamma$, so one may not set $G\to 0$ if one does not also set 
$\gamma\to 0$.  Therefore, the case with $G\ne 0$ may be more relevant.  
In this case 
$\langle \delta'|b,B,\Delta'\rangle
  = \gamma(1-G^2)(b - \langle b|B,\Delta'\rangle)$.
Since this same term appears in the exponential of 
$p(b|B,\Delta')$, we have that $f(s|B,\Delta')$ 
has the same form as equation~\eqref{sfs}, except for the shift
 $b\to b - \langle b|B,\Delta'\rangle$.  Therefore, except for the 
erf piece, the final integral over $\Delta'$ can also be done analytically.   
We leave a comparison of this approximation, and the full expression 
(in which we have not ignored $\langle\delta'\Delta\rangle$ etc.), 
with the numerical Monte-Carloed distributions, to future work.  

That $f(s|B,\Delta')$ depends explicitly on $\Delta'$, even in this 
approximation, is significant, since it shows that   
the conditional distribution for first crossing $b$ depends not 
just on the fact that $B$ was crossed, but on how $B$ was crossed.  
If we interpret `how $B$ was crossed' as a statement about the mass 
distribution on smoothing scales larger than $S$, then the fact 
that $f(s|B,\Delta')$ depends on $\Delta'$ indicates that the formation 
history of the mass within $S$ depends on the surrounding environment.  
This shows that our formalism will naturally give rise to 
`Assembly Bias' effects of the sort identified 
by Sheth \& Tormen (2004), and studied since by many others.

\subsection{Dependence of $B$ on parameters other than $\delta$}\label{dcep}
In triaxial collapse models, the barrier is a function of the 
values of the initial deformation tensor (rather than just its trace).  
This makes 
\begin{align}
 f(s)\,{\rm d}s = 
 &\int {\rm d}e \int_{-e}^e {\rm d}p \int {\rm d}e' \int {\rm d}p' 
  \int_{B'}^\infty {\rm d}\delta'\nonumber\\
 &\times \int_{\delta_c(e,p)}^{\delta_c(e,p) + \Delta s\,(\delta'-B')}
                     \!\!{\rm d}\delta\,p(e,p,\delta,e',p',\delta') 
\end{align}
where $B' = e'\, (\partial B/\partial e) + p'\, (\partial B/\partial p)$.  
If the scale dependence of $e$ and $p$ can be neglected, then this will 
simplify to 
\begin{align}
 f(s) = &\int {\rm d}e \int_{-e}^e {\rm d}p\, p(e,p,\delta_c(e,p)) \notag \\
 &\times \int_0^\infty {\rm d}\delta'\, \delta' \,p(\delta'|e,p,\delta_c(e,p)) .
\end{align}
Further, if the correlation between $\delta'$ and $e$ and $p$ can be 
neglected, then 
\begin{equation}
 f(s) = \int {\rm d}e \int_{-e}^e {\rm d}p\, g(e,p|\delta_c(e,p))\,
       f(s|\delta_c(e,p)),
\end{equation}
and the first crossing distribution is that for $\delta_c(e,p)$, 
weighted by the probability of having $e$ and $p$.  
If we use equation~(A3) of Sheth et al. (2001) for the distribution of 
$e$ and $p$ given $\delta$, then our analysis should be thought of as 
generalizing their equation~(9). 
In particular, because the result is a weighted sum of first crossing 
distributions, it exhibits exactly the sort of stochasticity discussed 
in Appendix C of Paranjape et al. (2011).  Performing this calculation 
more carefully is the subject of ongoing work.  

\subsection{Excursion set model of peaks}\label{peaks}
One of the great virtues of our approach is that it provides a simple 
way to compare the excursion set description with that for peaks.  
The relation is particularly simple for Gaussian smoothing filters, 
since then $\gamma$ of our equation~\eqref{gammas} is the {\em same} 
parameter which plays an important role in peaks theory.  This 
correspondence means that our parameter $x$ is essentially the same as 
the peak curvature parameter; the only difference is that we define the 
derivative with respect to the variance $s$, whereas peaks theory 
derivatives are with respect to smoothing scale $R$.  This means that 
our integrals over $\delta'$ are really just integrals over curvature, 
which makes intuitive sense.  Hence, to implement our prescription for 
peaks, we only need to account for the fact that the distribution of 
curvatures around a peak position differs from that around random 
positions.  If we write our equation~\eqref{likepeaks} as 
$(\beta_*/2)\,{\rm e}^{-\beta^2/2}/\sqrt{2\pi}$ 
times $\langle x\rangle/\gamma\beta_*$, then we need only replace 
$\langle x\rangle \to \langle xf(x)\rangle$, where $f(x)$ is given 
by equation~(A15) of Bardeen et al. (1986).  

For a constant barrier, our $\gamma\beta_*$ equals their $x_*$, so 
our expression is simply their equation~(A14) weighted by $x/x_*$ and 
integrated over $x$.  (Their additional factor of $(2\pi)^{3/2} R_*^3$ 
is just the usual $m/\bar\rho$ factor in excursion set theory, which 
converts from mass fractions to halo abundances.)   
Omitting the $x/x_*$ factor when performing the integral over $x$ 
yields the usual expression for peaks (their equation~A18).  
Therefore, one might think of their~(A18) as representing the 
`completely correlated limit' for peaks, whereas our analysis yields 
what should be thought of as the moving barrier excursion set model 
for peaks.   

\section{Discussion}
We presented a formula, equation~\eqref{sfs}, which provides an 
excellent description (see Fig.~\ref{compareMC}) of the first crossing 
distribution of a large variety of barriers, by walks exhibiting a 
large range of correlations (i.e., it is valid for a wide class of 
power spectra and smoothing filters).  As we discuss below, we expect 
a special case of it -- equation~\eqref{sfsconstant} -- to be a 
good physically motivated fitting formula for halo abundances.  

We then showed that our approach provides a simple expression for 
the first crossing distribution associated with walks conditioned to
pass through a certain point, and hence a simple expression for how 
halo bias factors are modified because of the correction term
(equation~\ref{bias1}).  
We sketched why a generic feature of this approach is that even the 
linear bias factor should be scale dependent.  
We also showed how to approximate the first crossing distributions 
associated with two non-intersecting barriers:  the probability 
that a walk first crosses barrier $B$ on some scale $S$ and then $b>B$ 
on scale $s>S$ (equation~\ref{sfsS}).  This exhibit `assembly bias' so 
they may provide useful approximations for halo progenitor 
mass functions and merger rates.  

Finally, we argued that our approach makes it particularly easy to 
see how the first crossing distribution is modified if the barrier 
depends on hidden parameters, such as those associated 
with the triaxial collapse model (Section~\ref{dcep}) or peaks 
(Section~\ref{peaks}).  In our approach, peaks differ from 
random positions only in that the integral over $x$ in 
equation~\eqref{likepeaks} is modified.  A similar integral in the 
expression for peak abundances leads to scale dependent bias even 
at the linear level (Desjacques et al. 2010), and is why we now expect 
$k$-dependent bias even for random positions. 

We are not the first to have considered the correlated steps problem.  
Peacock \& Heavens (1990) identified $\Gamma$ (our equation~\ref{gammas}) 
as the key parameter.  Their approximation for $f(s)$ is more accurate 
than more recent approximations (Maggiore \& Riotto 2010; 
Achitouv \& Corasaniti 2011) which are, in any case, restricted to 
special combinations of power spectra and smoothing filter 
(Paranjape et al. 2012).  However, our equation~\eqref{sfs} is simpler 
and more accurate for a wider range of barrier shapes, power spectra 
and smoothing filters than any of these previous studies, and it can be 
easily extended. Besides the problems outlined above, an obvious
direction would be to include non-Gaussianity, along the lines of Musso \& Paranjape (2012). This extension is conceptually straightforward -- equations \eqref{limfs} will hold 
also for a non-Gaussian field -- and is currently being investigated.

We argued that this accuracy stems from the small $s$ behavior of the walks.
In this regime, which is of most interest in cosmology, $f(s)$ 
is well-described by equation~\eqref{fcc} for completely correlated 
walks (Paranjape et al. 2012).  These walks are deterministic: 
the distribution of their heights is 
a delta function. For real walks, instead, it has a width that 
increases with $s$.  Equation~(\ref{sfs}) was obtained by allowing for this broader distribution of heights in the one step prior to 
the crossing.
However, it does not explicitly account for walks which may have 
criss-crossed the barrier more than once.  
Musso et al. (2012) discuss how accounting for more zigs and zags may 
yield even greater accuracy at larger $s$.  


Equation~(\ref{sfs}) is explicitly the completely correlated first crossing 
rate times a correction factor.  It depends on power spectrum only 
because this factor depends on $\Gamma$.  Since this factor is small at 
small $s$, in this regime (which is the usual one in cosmology) 
one should expect to see departures from self-similarity, but they should 
be small.  E.g., for a barrier of constant height, the first crossing 
distribution becomes equation~\eqref{sfsconstant}, and the correction 
factor becomes important at $\nu\le \Gamma^{-1}$, with $\Gamma^{-1}\sim 3$ 
for the $\Lambda$CDM family of $P(k)$.  

In cosmology the first crossing distribution is often used as a 
fitting formula.  For this purpose, one might treat either 
$\delta_c$ or $\Gamma$ or both as free parameters (even though 
equation~\ref{sfsconstant} describes the first crossing distribution 
very well with no free parameters).  In this case, the value of 
$\delta_c$ will depend on a variety of factors (Sheth et al. 2001; 
Maggiore \& Riotto 2010b; Paranjape et al. 2012), and we expect 
$\Gamma$ to depend on the effective slope of the power spectrum.  
In this sense, our formula provides a simple way to understand, 
interpret and quantify the departures from self-similarity which 
simulations are just beginning to show.  

\section*{Acknowledgments}
It is a pleasure to thank Aseem Paranjape for helpful comments on the draft.
RKS is supported in part by NSF-AST $0908241$.

\label{lastpage}

\end{document}